\documentclass[11pt, letterpaper]{article}
\usepackage{mydef2col}
\usepackage{kantlipsum,widetext}
\usepackage[T1]{fontenc}
\usepackage[autostyle=true]{csquotes}

\allowdisplaybreaks[3]
\graphicspath{{./Graphs/}}

\def\calF{{\cal F}}
\def\calV{{\cal V}}
\def\calG{{\cal G}}
\def\calVH{{\cal V}_H}
\def\calY{{\cal Y}_H}
\def\calx{{\chi}}
\def\Ito{It\^{o}'s }

\newcommand{\iu}{\mathrm{i}\mkern1mu}
\newcommand{\dom}[ 1 ]{\mathrm{Dom}(#1)}

\title{ADOL - Markovian approximation of rough lognormal model}
\author{
\authorstyle{Peter Carr
and Andrey Itkin{}
}
\newline\newline
\institution{Tandon School of Engineering, New York University, 12 Metro Tech Center, 26th floor, Brooklyn NY 11201, USA}}

\date{\today}
\begin{document}

\maketitle

\lettrineabstract{In this paper we apply Markovian approximation of the fractional Brownian motion (BM), known as the Dobric-Ojeda (DO) process, to the fractional stochastic volatility model where the instantaneous variance is modelled by a lognormal process with drift and fractional diffusion. Since the DO process is a semi-martingale, it can be represented as an \Ito diffusion. It turns  out that in this framework the process for the spot price $S_t$ is a geometric BM with stochastic instantaneous volatility $\sigma_t$, the process for $\sigma_t$ is also a geometric BM with stochastic speed of mean reversion and time-dependent colatility of volatility, and the supplementary process $\calV_t$ is the Ornstein-Uhlenbeck process with time-dependent coefficients, and is also a function of the Hurst exponent. We also introduce an adjusted DO process which provides a uniformly good approximation of the fractional BM for all Hurst exponents $H \in [0,1]$ but requires a complex measure. Finally, the cha-}
\begin{addmargin}[1cm]{1cm}
\small{\textbf{racteristic function (CF) of $\log S_t$ in our model can be found in closed form by using asymptotic expansion. Therefore, pricing options and variance swaps (by using a forward CF) can be done via FFT, which is much easier than  in rough volatility models.}}
\end{addmargin}

%%%%%%%%%%%%%%%%%%%%%%%%%%%%%%%%%%%%%%%%%%%%%%%%%%%%%%%%%%%%
\vspace{0.5in}

\section*{Introduction}

It was discovered in \cite{GatheralJaissonRos2014} that for a wide range of assets, historical volatility time-series exhibit a behavior which is much rougher than that of a Brownian motion. It was also shown that dynamics of log-volatility is well modeled by a fractional Brownian motion with Hurst parameter of order 0.1. Note that in the literature there exist various opinions whether the Hurst index should be less than 1/2 (short memory) or above 1/2 (long memory) depending on the particular asset class. As mentioned in \cite{Funahashi2017}, it is well known that i) the decrease in the market volatility smile amplitude is much slower than that predicted by the standard stochastic volatility models, and (ii) the term structure of the at-the-money volatility skew is well approximated by a power-law function with the exponent close to zero. These stylized facts cannot be captured by standard models, and while (i) has been explained by using a fractional volatility model with Hurst index $H > 1/2$, (ii) is proven to be satisfied by a rough volatility model with $H < 1/2$ under a risk-neutral measure. In more detail see \cite{Funahashi2017} and references therein.
In \cite{Livieri2018}  the value of the Hurst exponent obtained  by using high
frequency volatility estimations from historical price data have been revisited
by studying implied volatility based approximations of the spot volatility. Using at-the-money options on the S\&P500 index with short maturity, the authors confirm that volatility is rough, and the Hurst parameter is of order 0.3, i.e. slightly larger than that usually obtained from historical data.

Despite the fact that rough volatility models have been already well elaborated, and there exists rich literature on the subject, due to the non-Markovian nature of the fractional Brownian motion, one can face some technical problems when it comes to derivatives pricing. For instance, in \cite{EuchRos2016} using an original link between nearly unstable Hawkes processes and fractional volatility models, the authors compute the characteristic function of the log-price in rough Heston models. They show that in this case the Heston Riccati equation is replaced by a fractional Riccati equation which they solve numerically. Pricing variance swaps is even more complicated. Therefore, in this paper we make a try to attack the rough volatility problem by using some approximation to the Fractional Brownian motion which, however, is a semi-martingale.

\section{The DO process}

As mentioned, this paper aims to construct a rough lognormal model by replacing the fractional Brownian motion, driving the instantaneous volatility, with a similar process first introduced in \cite{DobricOjeda2009}. To provide a short description of this process, which further for the sake of brevity we call the DO process, below we follow \cite{DobricOjeda2009, Conus2016,Wildman2016}.

The DO process is a Gaussian Markov process with similar properties to those of a fractional Brownian motion, namely its increments are dependent in time. The DO process is defined by first considering the fractional Gaussian field
$Z = Z_H(t), \  (t,H)\in [0,\infty)\times (0,1)$ on a probability space $(\Omega, \cal F, \mathbb{P})$ defined by covariance (compare this with a standard fractional BM where $\alpha_{H,H'} = 1$, and $H = H'$)
\begin{widetext}
\begin{align}
\EE [Z_H(t) Z_{H'(t)}] &= \frac{\alpha_{H,H'}}{2}\left[|t|^{H+H'} + |s|^{H+H'} - |t-s|^{H+H'}\right], \\
\alpha_{H,H'} &=
\begin{cases}
-\dfrac{2 \eta}{\pi} \xi(H) \xi(H') \cos \left[\dfrac{\pi}{2}(H'-H)\right]
\cos \left[\dfrac{\pi}{2}(H'+H)\right], & H=H' \ne 1, \\
\xi \sin^2(\pi H) \equiv \alpha_h \equiv \alpha_{H'}, & H + H' = 1,
\end{cases}
\nonumber \\
\xi(H) &= \left[\Gamma(2H+1)\sin(\pi H)\right]^{1/2}, \quad
\eta = \Gamma(-(H+H')), \quad
\xi = \left[\Gamma(2H+1)\Gamma(3-2H)\right]^{1/2}. \nonumber
\end{align}
\end{widetext}
\noindent Here $\Gamma(x)$ is the Gamma function, \cite{as64}. Obviously, if $H=H'$, $Z_H$ is a fractional Brownian motion, and so if $H=H'=1/2$ it is a standard Brownian motion. It was established in \cite{DObook2006} that $Z_H$ exists.

Further \cite{DobricOjeda2009} are seeking for a process of the form $\psi_H(t) M_H(t)$ that in some sense approximates fractional Brownian motion, assuming that $\psi_H(t)$ is a deterministic function of time, and $M_H(t)$ is a stochastic process. They construct $M_H(t)$ as follows. On the Gaussian field $Z$ define $M_H(t), \ t \in [0,\infty)$ as
\begin{equation} \label{MH}
M_H(t) = \EE[Z_{H'}(t) | \calF_t^H],
\end{equation}
\noindent where $\calF_t^H$ is a filtration generated by a sigma-algebra $Z_H(r)), \ 0 \le r \le t$. It is proved in \cite{Conus2016,DobricOjeda2009}, that $M_H(t)$ is a martingale with respect to $(\calF_t^H)_{t \ge 0}$. It is also shown that $M_H(t)$ is  a Gaussian centered process with independent increments and covariance
\begin{align} \label{EMM}
\EE [M_H(t) M_H(s)] &= c_H \alpha_H \bar{B}(3/2-H)(s \wedge t)^{2-2 H}, \\
c_H &=  \dfrac{\alpha_H}{2 H \Gamma(3/2-H) \Gamma(H+1/2)}, \nonumber
\end{align}
\noindent where $\bar{B}(x) = B(x,x), \ B(x,y)$ is the Beta function.

The coefficient $\psi_H(t)$ could be determined  by minimizing the difference $\EE [(Z_H(t) - \psi_H(t) M_H(t))^2]$ to provide
\begin{equation} \label{psi}
\psi_H(t) = \dfrac{\EE [Z_H(t) M_H(t)]}{\EE [M^2_H(t)]},
\end{equation}
\noindent and, as shown in \cite{DobricOjeda2009}, in the closed form
\begin{equation}
\psi_H(t) = \dfrac{\Gamma(3-2H)}{c_H \Gamma^2(3/2-H)} t^{2H-1}.
\end{equation}

To summarize this construction, it introduces the DO process $V_H(t), \ t \in [0,\infty]$ defined as $V_H(t) = \psi_H(t) M_H(t)$ where $\psi_H(t)$ is given in \eqref{psi}, and $M_H(t)$ - in \eqref{MH} with $H+H'=1$.

The most useful property of the DO process is that it is a semi-martingale, and can be represented as an \Ito diffusion. This means, see again \cite{DobricOjeda2009,Wildman2016}, that there exists a Brownian motion process $W_t, \  t \in [0,\infty)$ adapted to the filtration $\calF_t^H$, such that
\begin{align} \label{DifRepr}
dV_H(t) &= \dfrac{2H-1}{t} V_H(t) dt + B_H t^{H-1/2} d W_t, \\
B_H &= \frac{2^{3-4 H} \csc ^4(\pi  H) \Gamma (2-H)}{\Gamma \left(3/2-H\right)^2 \Gamma (H)}. \nonumber
\end{align}

In contrast to \cite{Conus2016} where the DO process was used as noise in the Black-Scholes framework, here we apply it for modeling dynamics of the instantaneous variance. The main advantage of such a model as compared with the rough volatility models is that the semi-martingale property of the DO process allows utilization of the \Ito calculus.

Also in a recent paper \cite{Harms2019} it is shown that fractional Brownian motion can be represented as an integral over a family of the Ornstein-Uhlenbeck (OU) processes. The author proposes numerical discretizations which have
strong convergence rates of arbitrarily high polynomial order. He uses this representation as the basis of Monte Carlo schemes for fractional volatility models, e.g. the rough Bergomi model. Thus, the DO process can be considered as a particular case of the construction in \cite{Harms2019}. However, as we show below, using the DO approximation of the fractional Brownian motion provides some additional tractability, while is less accurate.

\section{An adjusted DO process}

As $\psi(t)$ in \eqref{psi} is determined by minimization of variance of the process $Y_H(t) = Z_H(t) - \psi_H(t) M_H(t)$, let us derive an explicit representation of this minimal value $\EE[Y^2_H(t) | \calF_t^H]$. As shown in \cite{DobricOjeda2009},
\begin{equation} \label{EZM}
\EE[Z_H(t) M_H(t)] = \EE[Z_H(t) Z_{H'}(t)] = \alpha_H t.
\end{equation}
Now, using \cref{EMM,psi,EZM} one can derive
\begin{align}
\EE[Y^2_H(t)] &= t^{2H} - \dfrac{\Big\{\EE [Z_H(t) M_H(t)]\Big\}^2}{\EE [M^2_H(t)]} = d_H^2 t^{2H} = d_H^2 \EE[Z^2_H(t)], \\
d_H^2 &= 1 - 2H \dfrac{\Gamma(3-2H) \Gamma(H+1/2)}{\Gamma(3/2-H) }. \nonumber
\end{align}

The last expression indicates that for $H \in [0.4,1]$ the process $V_H$ approximates $Z_H$ with a relative $L^2$ error at most at 12\%, see Fig.1 in \cite{DobricOjeda2009}. At lower $H$ the discrepancy is bigger and can reach 80-100\% at small $H$. However, based on the survey presented in Introduction, the Hurst exponent could vary for various markets, and the region $H < 0.4$ is important in practice.

From this prospective we introduce an adjusted DO (ADO) process which is defined as
\begin{equation} \label{ADO}
\calVH(t) = \psi_H(t) M_H(t)  + \iu d_H t^H = V_H(t) + \iu d_H t^H,
\end{equation}
\noindent with $i$ be an imaginary unit. The ADO process inherits a semi-martingale property from $V_H(t)$. Also $\psi_H(t)$ as it is defined in \eqref{psi}, still minimizes the difference $\EE[\calY^2(t)]  = \EE [(Z_H(t) - \calVH(t))^2]$. Finally, the minimum value of this difference is
\begin{equation}
\EE[\calY^2(t)] = \EE[\{Z_H(t) - (\psi_H(t) M_H(t) + \iu d_H t^H)\}^2] = 0.
\end{equation}
However, this requires an extension of the traditional measure theory into the complex domain, see, e.g., \cite{CarrWu2004}.

As from the definition, $V_H(t) = \calVH(t)  - \iu d_H t^H$, \eqref{DifRepr} can be transformed to
\begin{align} \label{DifRepr1}
d\calVH(t) &= \left[ \iu H d_H t^{H-1} + \dfrac{2H-1}{t} \calVH(t) \right] dt + B_H t^{H-1/2} d W_t, \end{align}
\noindent with the same Brownian motion as in \eqref{DifRepr}. In other words, the ADO process can also be represented as an \Ito diffusion. If $H < 1/2$ it exhibits mean-reversion.

It is worth to underline that the ADO process is not a martingale any more under
$\calF_t^H$ due to the adjustment made. However, as we use this process for modeling the instantaneous variance, it should not be a martingale. Hence, the only property we need is that the ADO process is a semi-martingale, and it can be represented as an \Ito diffusion in \eqref{DifRepr1}.

As mentioned in \cite{Conus2016}, the term $1/t$ in the drift of $\calVH(t)$ causes explosion of the DO process at $t=0$. To remedy this issue, they define a modified process, in which the drift is 0 until time $t = \epsilon > 0$. Here we exploit this idea for the ADO process as well.

\section{The ADOL model}

One of the most popular stochastic volatility (SV) models of \cite{Heston:93} introduces an instantaneous variance $v_t$ as a mean-reverting square-root process correlated to the underlying stock price process $S_t$. The model is defined by the following stochastic differential equations (SDEs):
\begin{align} \label{heston}
dS_t &= S_t(r-q) dt + S_t\sqrt{v}dW^{(1)}_t \\
d v_t &= \kappa (\theta - v_t)dt + \xi\sqrt{v_t} dW^{(2)}_t, \nonumber \\
S_t\big|_{t=0} &= S_0, \quad v_t\big|_{t=0} = v_0. \nonumber
\end{align}
\noindent where $W^{(1)}$ and $W^{(2)}$ are two correlated Brownian motions with the constant correlation coefficient $\rho$, $\kappa$ is the rate of mean-reversion, $\xi$ is the volatility of variance $v$ (vol-of-vol), $\theta$ is the mean-reversion level (the long-term run), $r$ is the interest rate and $q$ is the continuous dividend. All parameters in the Heston model are assumed to be time-independent, despite this assumption could be relaxed, \cite{Benhamou2010}.

As mentioned in Introduction, analysis of the market data reveals a rough nature of the implied volatility. Therefore, to take this into account in \cite{EuchRos2016} a fractional version of the Heston model was introduced. The authors consider the case $H \in [0, 1/2]$ where their rough Heston model  is neither Markovian, nor a semi-martingale. An alternative rough Heston models is proposed in \cite{GueJacRoome2014}. The main result obtained in \cite{EuchRos2016} is that the characteristic function of the log-price in rough Heston models exhibits the same structure as that one in the classical Heston model. However, the corresponding Riccati equation, see eg, \cite{Rouah2013},  is replaced by a fractional Riccati equation. This equation doesn't have an explicit  solution anymore, but can be solved numerically by transforming it to some Volterra equation.

In addition to the Heston model, a similar model but written in terms of the volatility, rather than variance, was also given some attention  in the literature. The model is defined by the following SDE:
\begin{align} \label{lognorm}
dS_t &= S_t(r-q) dt + S_t \sigma_t dW^{(1)}_t \\
d \sigma_t &= \kappa (\theta - \sigma_t)dt + \xi \sigma_t dW^{(2)}_t, \nonumber \\
S_t\big|_{t=0} &= S_0, \quad \sigma_t\big|_{t=0} = \sigma_0. \nonumber
\end{align}
Thus, it is a mean-reverting lognormal model for the instantaneous volatility $\sigma_t$. This model is a flavor of a famous SABR model of \cite{hagan2002}, and is also advocated, e.g., in \cite{Sepp2016}. The latter paper claims that working with the model dynamics for $\sigma_t$ is more intuitive, and provides a clearer interpretation of the parameters in terms of the log-normal SABR model, which is well understood by practitioners. As far as the market data is concerned, \cite{Sepp2016} makes a reference to \cite{CJM2010} who examined the empirical performance of the Heston, lognormal and 3/2 SV models using market data on VIX, the implied volatility of S\&P500 options, and the realized volatility of S\&P500 returns. It was found that the lognormal model outperforms the others. For more discussion, again see \cite{Sepp2016}.

As far as pricing options under the lognormal SV model is concerned, in contrast to the Heston model, the former is not affine. Therefore, yet a closed-form solution for the characteristic function for the $\log S_T$ is not known, while some approximations were reported in the literature. In \cite{Lewis:2000} this characteristic function is constructed assuming $\theta = 0$ and by using a Hypergeometric series expansion. In \cite{Sepp2016} an approximate solution is constructed by using a series expansion in the centered volatility process $Y_t = \sigma_t - \theta$.

The main idea of this paper, however, is to propose a tractable version of the rough lognormal model. For doing that we achieve the following steps:
\begin{enumerate}
\item For the instantaneous volatility process instead of the fractional Brownian motion we use the ADO process.

\item Similar to \cite{Lewis:2000} we assume the mean-reversion level $\theta=0$. This, however, can be relaxed, see discussion at the end of this paper.

\end{enumerate}
Also, further for simplicity of notation we will use symbols $\calV_t$ instead of $\calVH(t)$, and $\nu(t) = B_H t^{H-1/2}$ . Then, assuming real-world dynamics (i.e., under measure $\mathbb{P}$), our model could be represented as
\begin{align} \label{ADOheston3}
& dS_t = S_t\mu dt + S_t\sigma_t dW^{(1)}_t \\
& d \sigma_t = \sigma_t \left[- \kappa + \xi D_v \right] dt + \sigma_t \xi \nu(t) dW^{(2)}_t,  \nonumber \\
& d\calV_t = D_v dt + \nu(t) dW^{(2)}_t , & \nonumber \\
& D_v = \left[ \iu H d_H t^{H-1} + \dfrac{2H-1}{t} \calV_t \right] \ST{t > \epsilon}, \nonumber \\
S_t\big|_{t=0} &= S_0, \quad \sigma_t\big|_{t=0} = \sigma_0,
\quad \calV_t\big|_{t=0} = \calV_0, \nonumber
\end{align}
\noindent where $\mu$ is the drift. This model is a two-factor model (actually, we introduced three stochastic variables $S_t, \sigma_t, \calV_t$, but two of them: $\sigma_t$ and $\calV_t$ are fully correlated).

The model in \eqref{ADOheston3} is a stochastic volatility model where the speed of mean-reversion of the instantaneous volatility $\sigma_t$ is stochastic, but fully correlated with $\sigma_t$. In the literature there have been already some attempts to consider an extension of the Heston model by assuming the mean-reversion level $\theta$ to be stochastic, see \cite{Gatheral2008,Bi2016}. In particular, in \cite{Gatheral2008} it is shown that such a model is able to replicate a term structure of VIX options. However, to the best of our knowledge, stochastic mean-reversion speed has not been considered yet. In what follows, for the sake of brevity we call it ADOL - the adjusted DO lognormal model. Also in our model the vol-of-vol is time-dependent.

To use this model for option pricing, the stock price $S_t$ should be a martingale under the risk-neutral measure $\mathbb{Q}$. Then, for instance, for the Heston model an additional restriction was proposed in \cite{Heston:93} that the market price of volatility risk is $\lambda \sqrt{v_t}$, where $\lambda = const$. This is dictated by tractability (while a financial argument is also available, see \cite{WongHeyde2006} and references therein), because then the SDE for $v_t$ has the same functional form under $\mathbb{P}$ and $\mathbb{Q}$ assuming both measures exist.

It can be seen from \eqref{ADOheston3} that under the ADOL model  the process for $S_t$ is a geometric Brownian motion with stochastic instantaneous volatility $\sigma_t$, the process for $\sigma_t$ is also a geometric Brownian motion with stochastic speed of mean reversion and time-dependent vol-of-vol, and the process for $\calV_t$ is the (OU) process with time-dependent coefficients. As by definition in \eqref{ADOheston3} the drift $D_v$ vanishes at $t=0$, the mean-reversion speed of $\sigma_t$ at the origin becomes $-k$, i.e. is well-defined $\forall H \in [0,1]$.

\section{The ADOL partial differential equation (PDE)}

To price options written on the underlying stock price $S_t$ which follows the ADOL model, a standard approach can be utilized, \cite{Gatheral2006,Rouah2013}. Consider a portfolio consisting of one option $V = V (S, \sigma, \calV, t)$, $\Delta$ units of the stock $S$, and $\phi$ units of another option $U = U(S, \sigma, \calV, t)$ that is used to hedge the volatility. The dollar value of this portfolio is
\begin{equation} \label{port}
\Pi = V + \Delta S + \phi U.
\end{equation}
The change in the portfolio value $d \Pi$ could be found by applying \Ito  lemma to $dV$ and $dU$, and assuming that the continuous dividends are re-invested back to the portfolio
\begin{align} \label{port2}
d\Pi &= d V + \Delta d S + \phi d U + \Delta q S dt, \\
&= \Bigg\{ \fp{V}{t} + \frac{1}{2} \sigma^2 S^2 \sop{V}{S} + \frac{1}{2} \xi^2 \sigma^2 \nu^2(t) \sop{V}{\sigma} + \frac{1}{2} \nu^2(t) \sop{V}{\calV} \nonumber \\
&\hspace{0.55in} +  \rho S \xi \sigma^2  \nu(t) \cp{V}{S}{\sigma} + \rho S \sigma \nu(t) \cp{V}{S}{\calV} + \xi \sigma \nu^2(t) \cp{V}{\calV}{\sigma} \Bigg\} dt \nonumber \\
&+ \phi \Bigg\{ \fp{U}{t} + \frac{1}{2} \sigma^2 S^2 \sop{U}{S} + \frac{1}{2} \xi^2 \sigma^2 \nu^2(t) \sop{U}{\sigma} + \frac{1}{2} \nu^2(t) \sop{U}{\calV} \nonumber \\
&\hspace{0.55in} + \rho S \xi \sigma^2  \nu(t) \cp{U}{S}{\sigma} + \rho S \sigma \nu(t) \cp{U}{S}{\calV} + \xi \sigma \nu^2(t) \cp{U}{\calV}{\sigma} \Bigg\} dt \nonumber \\
&+ \Bigg\{ \fp{V}{S} + \phi \fp{U}{S} + \Delta \Bigg\} dS + \Bigg\{ \fp{V}{\sigma} + \phi \fp{U}{\sigma} \Bigg\} d\sigma +
\Bigg\{ \fp{V}{\calV} + \phi \fp{U}{\calV} \Bigg\} d \calV + \Delta q S dt. \nonumber
\end{align}
Based on \eqref{ADOheston3}, the last three terms in \eqref{port2} in the explicit form could be re-written as
\begin{align} \label{port1}
\Bigg\{ \fp{V}{S} &+ \phi \fp{U}{S} + \Delta \Bigg\} dS + \Bigg\{ \fp{V}{\sigma} + \phi \fp{U}{\sigma} \Bigg\} d\sigma +
\Bigg\{ \fp{V}{\calV} + \phi \fp{U}{\calV} \Bigg\} d \calV \\
&= \Bigg\{ \fp{V}{S} + \phi \fp{U}{S} + \Delta \Bigg\}\left[S \mu dt + S \sigma d W^Q_{1,t}\right]
\nonumber \\
&+ \Bigg\{ \fp{V}{\sigma} + \phi \fp{U}{\sigma} \Bigg\}\sigma\left[-\kappa + \xi \bar{D}_v \right] dt +
\Bigg\{ \fp{V}{\calV} + \phi \fp{U}{\calV} \Bigg\} \bar{D}_v dt \nonumber \\
&+   \nu(t) d W^Q_{2,t} \Bigg\{ \left[\fp{V}{\calV} + \phi \fp{U}{\calV} \right] + \xi \sigma \left[ \fp{V}{\sigma} + \phi \fp{U}{\sigma}\right] \Bigg\}. \nonumber
\end{align}

To make this portfolio riskless, the risky terms proportional to increments of the Brownian Motions must vanish. This implies that the hedge parameters are
\begin{align} \label{pars}
\Delta &= -\fp{V}{S} - \phi \fp{U}{S}, \\
\phi &= -\left[\xi \sigma \fp{V}{\sigma} + \fp{V}{\calV} \right]\left[ \xi \sigma  \fp{U}{\sigma} +  \fp{U}{\calV} \right]^{-1}. \nonumber
\end{align}
Also a relative change of the risk free portfolio is the interest earned with the risk free interest rate, i.e.
\begin{equation} \label{rn}
d \Pi = r \Pi dt.
\end{equation}
With allowance for \eqref{pars}, \eqref{port2} could be represented in the form $d\Pi = (A + \phi B) dt$. Therefore, \eqref{rn} can be transformed to
\begin{equation} \label{rn1}
A + \phi B  = r (V + \Delta S + \phi U).
\end{equation}
Using the definition of $\phi$ in \eqref{pars}, this could be re-written as
\begin{equation} \label{rn3}
\frac{A - r V + (r-q) S \fp{V}{S} }{\xi \sigma \fp{V}{\sigma} + \fp{V}{\calV} }  = \frac{B - r U + (r-q) S \fp{U}{S}}{\xi \sigma  \fp{U}{\sigma} +  \fp{U}{\calV} }.
\end{equation}
The left-hand side of this equation is a function of $V$ only, and the right-hand side is a function of $U$ only. This could be only if both sides are just some function $f(S,v,\calV,t)$ of the independent variables. Accordingly, using the explicit expression for $A$ from \eqref{rn3} we obtain the ADOL PDE
\begin{align} \label{pde}
0 &= \fp{V}{t} + \frac{1}{2} \sigma^2 S^2 \sop{V}{S} + \frac{1}{2} \xi^2 \sigma^2  \nu^2(t) \sop{V}{\sigma} + \frac{1}{2} \nu^2(t)  \sop{V}{\calV} \\
&+  \rho S \xi \sigma^2  \nu(t) \cp{V}{S}{\sigma} + \rho S \sigma \nu(t) \cp{V}{S}{\calV} + \xi \sigma \nu^2(t) \cp{V}{\calV}{\sigma}   \nonumber \\
&+ (r-q) S \fp{V}{S}  + (\bar{D}_v - f) \fp{V}{\calV} +
\sigma\left[-\kappa + \xi (\bar{D}_v - f) \right] \fp{V}{\sigma} - r V. \nonumber
\end{align}
To proceed, we need to choose an explicit form of $f(S,v,\calV,t)$. We consider two options.
The first one relies on a tractability argument and suggests to choose $f = \bar{D}_v + \lambda$, where, similar to \cite{Heston:93}, $\lambda$ is the market price of volatility risk and is constant. However, with this choice the risk-neutral drift of $\sigma_t$ becomes $- (\kappa + \xi \lambda)\sigma_t dt$, i. e., the stochastic volatility $\sigma_t$ doesn't depend on $\calV_t$. In such a model only the vol-of-vol term is a function of $t$ and the Hurst exponent $H$, so this is a stochastic volatility model with the time-dependent vol-of-vol. This makes this model not rich enough for our purposes, despite it is tractable. Therefore, in what follows we ignore this choice. For the reference, pricing options using the time-dependent Heston model is considered in \cite{Benhamou2010} by using an asymptotic expansion of the PDE in a small vol-of-vol parameter, and a similar method could be applied in this case as well.

The other construction we introduce in this paper is the choice $f = \bar{D}_v + \lambda + m(t) \calV$ with $m(t)$ be some function of time $t$. Since in \eqref{ADOheston3} the drift of $\sigma_t$ is already a linear function of $\calV_t$ under a physical measure, the proposed construction either keeps it linear under the risk-neutral measure.  With this definition \eqref{pde} takes the form
\begin{align} \label{pde1}
0 &=  \fp{V}{t} + \frac{1}{2} \sigma^2 S^2 \sop{V}{S} + \frac{1}{2} \xi^2 \sigma^2  \nu^2(t) \sop{V}{\sigma} + \frac{1}{2} \nu^2(t) \sop{V}{\calV}  +  \rho S \xi \sigma^2   \nu(t) \cp{V}{S}{\sigma} \\
&+ \rho S \sigma \nu(t) \cp{V}{S}{\calV} + \xi \sigma \nu^2(t) \cp{V}{\calV}{\sigma} + (r-q) S \fp{V}{S}  - [\lambda + m(t)\calV]\fp{V}{\calV} - [\kappa  + \xi (\lambda+m(t) \calV) ]\sigma  \fp{V}{\sigma} - r V. \nonumber
\end{align}

As by Girsanov's theorem, \cite{karatzas1991brownian}
\begin{align}
d W^{(1)}_t &= d W_{1,t}^{Q} - \gamma_1(t) dt, \\
d W^{(2)}_t &= d W_{2,t}^{Q} - \gamma_2(t) dt, \nonumber
\end{align}
\noindent with $W^Q, W_2^Q$ be the corresponding Brownian motions under measure $\mathbb{Q}$, a necessary condition for this measure to exist is
\[ \mu - (r-q) = \sigma_t \left( \rho \gamma_2(t) + \sqrt{(1-\rho^2} \gamma_1(t)\right),
\]
\noindent which ensures that the discounted stock price is a local martingale under measure $\mathbb{Q}$, see eg, \cite{WongHeyde2006}. Accordingly, by using the same argument, one can see that the PDE in \eqref{pde1} corresponds to the following model under the risk-neutral measure $\mathbb{Q}$
\begin{align} \label{ADOhestonQ}
dS_t &= S_t (r-q) dt + S_t\sigma_t dW^Q_{1,t} \\
d \sigma_t &= -  [\kappa  + \xi(\lambda+m(t)\calV_t) ] \sigma_t dt + \sigma_t \xi \nu(t) dW^{Q}_{2,t},  \nonumber \\
d\calV_t &= -[\lambda  + m(t)\calV]d t + \nu(t) dW^Q_{2,t} , & \nonumber \\
S_t\big|_{t=0} &= S_0, \quad \sigma_t\big|_{t=0} = \sigma_0, \quad \calV_t\big|_{t=0} = \calV_0. \nonumber
\end{align}
When this model is used for option pricing, and with parameters obtained by calibration of the model to market options prices, one is already in the risk-neutral setting. Then, as explained in \cite{Gatheral2006}, that allows setting the market price of volatility risk $\lambda$ equal to zero. So in what follows we set $\lambda = 0$.

The model for $\sigma_t$ in \eqref{ADOhestonQ} in a certain sense is similar to that introduced in \cite{StochMR2016} who considered a generalized OU process by letting a mean-reversion speed to be stochastic, and, in particular, a Brownian stationary process. As our process $\calV_t$ is also a time-dependent OU process, it may attain negative values, so the mean-reversion rate could become negative. However, in \cite{StochMR2016}, the authors are able to show the stationarity of the mean, the variance, and the covariance of the process (the process $\sigma_t$ in our notation) when the average speed of mean-reversion is sufficiently larger than its variance. Explicit conditions for these results to hold are also derived in that paper.

\section{Characteristic function of $\log S_T$ under the ADOL model} \label{CFT}

One of the main reasons that the Heston model is so popular is that the characteristic function of $\log S_T$ in this model is know in closed form. Then any FFT based method, \cite{CarrMadan:99a, Lewis:2000, FangOosterlee2008}, can be used to price European, and even American, \cite{LordAmerican2007}, options written on the underlying stock $S_t$.

Let us  denote $T$ to be the option maturity, and use the representation of the characteristic function  $\EE[e^{i u \log S_T} | S, v, \calV]  = e^{i u \log S} \psi(u; x,\sigma,\calV, t)$, where $\psi(u; x,\sigma,\calV, \tau) = \EE[e^{i u \log x}]$ and $x = \log S_T/S$.  It is known that as per Feynman-Kac theorem, \cite{Shreve:1992}, $\psi(u; x, \sigma,\calV,\tau)$ solves a PDE similar to \eqref{pde1} but with no discounting term $r V$
\begin{align} \label{pde11}
0 &=  \fp{\psi}{t} +   \frac{1}{2} \sigma^2 \sop{\psi}{x} + \frac{1}{2} \xi^2 \sigma^2  \nu^2(t) \sop{\psi}{\sigma} + \frac{1}{2} \nu^2(t) \sop{\psi}{\calV}  +  \rho \xi \sigma^2 \nu(t) \cp{\psi}{x}{\sigma} \\
&+ \rho \sigma \nu(t) \cp{\psi}{x}{\calV} + \xi \sigma \nu^2(t) \cp{\psi}{\calV}{\sigma} + \left(r - q -  \frac{1}{2}\sigma^2\right) \fp{\psi}{x}  - m(t)\calV \fp{\psi}{\calV} - (\kappa  + \xi m(t) \calV )\sigma \fp{\psi}{\sigma}, \nonumber
\end{align}
\noindent subject to the initial condition $\psi(u; x, \sigma,\calV, T) = 1$.

We will search the solution of this PDE in the form
\begin{equation} \label{subst0}
\psi(u; x, \sigma,\calV,t) = e^{\iu u x} z(u; t, \sigma, \calV),
\end{equation}
\noindent where $z(u; t, \sigma, \calV)$ is a new dependent variable. Substituting \eqref{subst0} into \eqref{pde11} yields
\begin{align} \label{pdePsi1}
0 &=  \fp{z}{t} + \frac{1}{2} \xi^2 \nu^2(t) \sigma^2 \sop{z}{\sigma}
+ \frac{1}{2} \nu^2(t) \sop{z}{\calV} + \xi \nu^2(t) \sigma \cp{z}{\calV}{\sigma} \\
&+ \left[- (\kappa  + \xi m(t) \calV) + \iu u \rho \xi \nu(t) \sigma\right] \sigma \fp{z}{\sigma}
+ \left[\iu u \rho \nu(t) \sigma - m(t) \calV\right] \fp{z}{\calV} +
\left[ -\frac{1}{2} u (\iu + u) \sigma^2 + \iu u (r - q)   \right] z, \nonumber
 \end{align}
 \noindent which should be solved subject to the initial condition $z(u; T, \sigma, \calV) = 1$.

To the best of our knowledge this PDE doesn't have a closed form solution. However, an approximate solution can be constructed. In particular, in what follows  we assume the vol-of-vol parameter $\xi$ to be small. More rigorously, observe that in the second line of \eqref{ADOhestonQ} the term $\Psi = \xi \nu(t) dW^{Q}_{2,t}$ is dimensionless. As $dW^{Q}_{2,t} \propto 1/(2\sqrt{t})$,  and $\nu(t) = B_H t^{H-1/2}$, we have $\Psi \propto \xi B_H t^H/2 = (\xi B_H T^H/2) (t/T)^H$. Suppose we consider only time intervals $0 \le t \le T$, hence
$0 \le (t/T)^H \le 1$. Then our assumption on $\xi$ being small means that $\xi B_H T^H/2 \ll 1$, or
\begin{equation} \label{smallVV}
\xi \ll \frac{2}{B_H T^H}.
\end{equation}
Obviously, this condition is too strong when we consider time intervals $t \ll T $, because then $(t/T)^H$ is also small. However, for relatively small maturities and small $H$ the latter could be violated even for small $t$. Therefore, we prefer not to rely on the smallness of $(t/T)^H$ even if it does take place, and consider \eqref{smallVV} as the definition of the small parameter.

With allowance for this assumption we construct the solution of \eqref{subst0} as follows. Let us
represent the solution of  \eqref{subst0} as a series
\begin{equation} \label{series1}
z(u;t,\sigma,\calV) = \sum_{i=0}^\infty \xi^i z_i(u;t,\sigma,\calV),
\end{equation}
\noindent where $\xi$ is a small parameter in a sense of \eqref{smallVV}. Substituting this representation into \eqref{subst0} yields

\begin{align} \label{pdeSeries}
0 &=  \sum_{i=0}^\infty \xi^i\fp{z_i}{t} + \sum_{i=0}^\infty \xi^i {\cal L} z_i  \\
&+ \frac{1}{2} \sum_{i=0}^\infty \xi^{i+2}  \nu^2(t) \sigma^2 \sop{z_i}{\sigma}
+ \sum_{i=0}^\infty \xi^{i+1} \left[\nu^2(t) \sigma \cp{z_i}{\calV}{\sigma}
+ \left(m(t) \calV + \iu u \rho \nu(t) \sigma\right) \sigma \fp{z_i}{\sigma} \right],
\nonumber \\
{\cal L} &=  \frac{1}{2} \nu^2(t) \sop{}{\calV}  - \kappa \sigma \fp{}{\sigma} +
\left[\iu u \rho \nu(t) \sigma - m(t) \calV\right] \fp{}{\calV} + \left[ -\frac{1}{2} u (\iu + u) \sigma^2 + \iu u (r - q)   \right]. \nonumber
\end{align}

It is clear that terms in the second line of \eqref{pdeSeries} have a higher order in $\xi$, and as such don't contribute, e.g., into the zero order solution. But for higher order approximations they appear  as source terms. In other words, the terms in the second line have no influence on Green's function of \eqref{pdeSeries}. This fact makes finding  the solution of \eqref{pdeSeries} much easier.

\subsection{Zero order solution of \eqref{pdeSeries}} \label{zeroOrder}

In the zero order approximation on $\xi$ \eqref{pdeSeries} transforms to
\begin{equation} \label{pdeSeries0}
0 =  \fp{z_0}{t} + {\cal L} z_0 .
\end{equation}
This equation could be solved in a few steps. First, we make a change of the dependent variable
\begin{align} \label{st1}
z_0(u;t,\sigma,\calV) &\mapsto y_0(u;t,\sigma,\calV) \exp\left[a(t) + \gamma(t) \sigma^2 + \beta(t) \sigma \calV \right], \\
\alpha(t) &= - \iu u (r-q) (t-T), \quad
\beta(t) = -\iu \rho  u \frac{1}{\nu(t)}, \quad
\gamma(t) = -\frac{u[1 + u (1 - \rho)^2]}{4 \kappa} \left(1 - e^{2 \kappa (t-T)} \right). \nonumber
\end{align}
With the new variable $y_0(u;t,\sigma,\calV)$ \eqref{pdeSeries0} transforms to
\begin{equation} \label{pdeSeries0-1}
0 = \fp{y_0}{t} + \frac{1}{2} \nu^2(t) \sop{y_0}{\calV} -
 m(t) \calV \fp{y_0}{\calV} - \kappa \sigma \fp{y_0}{\sigma} + \iu \rho \sigma
 \calV \frac{\nu'(t) + \nu(t)[\kappa + m(t)]}{\nu(t)^2} y_0,
\end{equation}
\noindent and should be solved subject to the initial (terminal) condition
$y_0(u;T,\sigma,\calV) = e^{-\beta(T) \sigma \calV}$.

Second, we introduce a new independent variable $\sigma \mapsto g = \sigma \calV$, and also will search the solution for the dependent variables
$y_0(u;t,g,\calV)$ in the form
\begin{equation} \label{prod}
y_0(u;t,g,\calV) = Y_1(u;t,g) Y_2(u;t,\calV).
\end{equation}
It turns out that after some algebra \eqref{pdeSeries0-1} in the new variables could be represented in the form
\begin{equation} \label{pdeSeries0-2}
\frac{1}{Y_1}\fp{Y_1}{t} - \frac{\kappa g }{Y_1}\fp{Y_1}{g} +
\iu \rho g \frac{\nu'(t) + \nu(t)[\kappa + m(t)]}{\nu(t)^2} =
-\frac{1}{Y_2}\fp{Y_2}{t} + m(t) \calV \frac{1}{Y_2}\fp{Y_2}{\calV} -
\frac{n(t)^2}{2 Y_2}\sop{Y_2}{\calV}.
\end{equation}
This equation has to be solve subject to the terminal condition $Y_1(t,g)Y_2(t,\calV) = e^{-\beta(T) g}$. Hence, we may impose the independent terminal conditions for $Y_1$ and $Y_2$ as
\begin{equation} \label{bcY}
Y_1(u;T,g) = e^{-\beta(T) g}, \qquad Y_2(u;T,\calV) = 1.
\end{equation}
A standard approach tells that since the LHS of \eqref{pdeSeries0-2} is a function of $(t,g)$ only, and the RHS of \eqref{pdeSeries0-2} is a function of $(t,\calV)$ only, both parts must be a function of $t$ only. In our case we can choose this function to be zero. This splits \eqref{pdeSeries0-2} into two independent equations
\begin{alignat}{2} \label{sysY}
\fp{Y_1}{t} &= \kappa g \fp{Y_1}{g} - \iu \rho g \frac{\nu'(t) + \nu(t)[\kappa + m(t)]}{\nu(t)^2} Y_1, \qquad  &&Y_1(u;T,g) = e^{-\beta(T) g}, \\
\fp{Y_2}{t} &= m(t) \calV \fp{Y_2}{\calV} -  \frac{1}{2} \nu(t)^2\sop{Y_2}{\calV}, \qquad &&Y_2(u;T,\calV) = 1. \nonumber
\end{alignat}
The first equation in \eqref{sysY} is a first order PDE (of the hyperbolic type),
and it can be easily solved in closed form to get
\begin{equation} \label{sysY1sol}
Y_1(u;t,g) = \exp \left[-\beta(T) g e^{k (t-T)} - \iu g \rho u \int_T^t \frac{\nu(t) [k+m(t)] + \nu'(t)}{\nu(t)^2} \, dt\right].
\end{equation}
The second equation is a convection-diffusion PDE of the type 3.8.7.4 in \cite{Polyanin2002} which can be reduced to the Heat equation. For instance, this can be done by doing a change of independent variables
\begin{align} \label{Htrans}
Y_2(u;t,\calV) &= e^{\alpha_1(t) \calV + \tau(t)} w(\tau, \varsigma), \\
\alpha_1(t) = e^{\int_T^t m(t) \, dt}, \qquad
\tau(t) &= - \frac{1}{2} \int_T^t \nu^2(s) \alpha_1^2(s) \, ds, \qquad
\varsigma = \alpha_1(t) \calV + 2 \tau(t), \nonumber.
\end{align}
In particualr, by this transformation the terminal point $t=T$ is mapped to $\tau = 0$. However, due to the terminal condition $Y_2(u;T,\calV) = 1$ the solution is just a constant $Y_2(u;t,\calV) = 1, \forall t \in [0,T]$.

Thus, combining all the above expressions into \eqref{st1}, we obtain
\begin{align} \label{st1final}
z_0(u;t,\sigma,\calV) &= \exp\left[\alpha(t) + \gamma(t) \sigma^2 + \bar{\beta}(t) \sigma \calV \right], \\
\bar{\beta}(t) &= \beta(t) - \beta(T) e^{k (t-T)} - \iu \rho u \int_T^t \frac{\nu(t) [k+m(t)] + \nu'(t)}{\nu(t)^2} \, dt \nonumber \\
&= \iu \rho u\left[\frac{e^{k (t-T)}}{\nu(T)} -
\frac{1}{\nu(t)} - \int_T^t \frac{k+m(t)}{\nu(t)} \, dt \right]. \nonumber
\end{align}

With this expression, the final representation of the CF in \eqref{subst0} reads
\begin{align} \label{finalCF0}
\psi(u; x, \sigma,\calV,t)\Big|_{t=0} &= \exp\left[\iu u x + a(0) + \gamma(0) \sigma^2 + \bar{\beta}(0) \sigma \calV \right], \\
a(0) &=  \iu u (r-q) T, \quad
\gamma(0) = -\frac{u[1 + u (1 - \rho)^2]}{4 \kappa} \left(1 - e^{-2 \kappa T} \right), \nonumber \\
\bar{\beta}(0) &= \iu \rho u\left[\frac{e^{-k T}}{\nu(T)} -
\frac{1}{\nu(0)} + \int_0^T \frac{k+m(t)}{\nu(t)} \, dt \right]. \nonumber
\end{align}
As by definition $\nu(t) = B_H t^{H-1/2}$, the expression for $\bar{\beta}(0)$ is well-defined only for $H < 1/2$. Then $1/\nu(0) = 0$.

\subsection{Green's function of the homogeneous PDE \eqref{pdeSeries0}} \label{GRF}

To construct higher order approximations in $\xi$ we need to determine Green's function of \eqref{pdeSeries0}. In our setting the Green function $\calG(\sigma,\sigma',\calV,\calV',t)$ should vanish at the boundary of the domain $\dom{\sigma} \dom{\calV} = [0,\infty] \times [-\infty,\infty]$ , and at $t=T$ it should be $\calG(\sigma,\sigma',\calV,\calV',T) = \Delta(\sigma - \sigma') \Delta(\calV - \calV')$.

The key point in determining the Green function of \eqref{pdeSeries0} is the representation \eqref{prod}.
Here we further modify it by making a change of variables
\begin{equation} \label{prod1}
y_0(u;t,g,\calV) =  e^{\alpha_1(t) \calV + \tau(t)} w_1(u;\tau, \omega) w_2(u;\tau, \varsigma), \qquad
\omega = e^{\kappa t} g,
\end{equation}
\noindent and $\tau(t), \alpha_1(t), \varsigma$ are defined in \eqref{Htrans}.

Accordingly, in new variables \eqref{pdeSeries0-2} takes the form
\begin{align} \label{wEq}
- a_1(t) \omega  &+ \dfrac{\partial_\tau w_1(u;\tau, \omega)}{w_1(u;\tau, \omega)} = \dfrac{-\partial_\tau w_2(u;\tau, \varsigma) + \partial^2_\omega w_2(u;\tau, \varsigma)}{w_2(u;\tau, \varsigma)}, \\
a_1(t) &= 2 \iu \rho  u \dfrac{\nu(t) (\kappa + m(t)) + \nu'(t)}{ \nu(t)^4}  e^{-\kappa t -2 \int_T^t m(t) \, dt} , \nonumber
\end{align}
\noindent and $t=t(\tau)$. This dependence can be obtained as an inverse of $\tau(t)$ defined in \eqref{Htrans}.

Again, the LHS of \eqref{wEq} is a function of $(\tau, \omega)$ only, while the RHS is a function of $(\tau, \varsigma)$ only. Therefore, both sides could be only some function of $\tau, i.e., f(\tau)$. In our case  we can put $f(\tau) = 0$.

As based on \eqref{prod1}, the solution of the PDE for $e^{-\alpha_1(t) \calV - \tau(t)} y_0(u;t,g,\calV)$ can be represented as a product $w_1(u;\tau, \omega) w_2(u;\tau, \varsigma)$, the Green function $\calG(\omega, \omega',\varsigma,\varsigma', \tau)$ can also be factorized, so
\begin{equation} \label{Gfact}
\calG(\omega, \omega',\varsigma,\varsigma', \tau) = \calG_1(\omega, \omega', \tau)\calG_2(\varsigma,\varsigma', \tau).
\end{equation}
The function $\calG_2(\varsigma,\varsigma',\tau)$ is the Green function of the Heat equation
\begin{equation}
\fp{w_2(u;\tau, \varsigma)}{\tau} = \sop{w_2(u;\tau, \varsigma)}{\omega},
\qquad \varsigma \in (-\infty, \infty),
\end{equation}
\noindent with $\calG_2(\varsigma,\varsigma',0) = \delta(\varsigma - \varsigma')$.  It is well-known and reads, \cite{Polyanin2002}
\begin{equation} \label{w2G}
 \calG_2(\varsigma,\varsigma',\tau) = \frac{1}{2\sqrt{\pi \tau}} e^{-\frac{(\varsigma - \varsigma')^2}{4 \tau}}.
\end{equation}

For the second equation
\begin{equation} \label{w1Eq}
\fp{w_1(u;\tau, \omega)}{\tau} =  a_1(t(\tau)) \omega w_1(u;\tau, \omega), \qquad   \omega \in (-\infty, \infty),
\end{equation}
\noindent the Green function can be found directly to obtain
\begin{equation} \label{w1G}
 \calG_1(\omega,\omega',\tau) = e^{\omega\int_0^\tau a(t(k)) dk} \delta(\omega - \omega')
 \left[ 1 - \Theta(-\tau) + \Theta(0) \right],
\end{equation}
\noindent where $\Theta(\tau)$ is the Heaviside theta-function, \cite{as64}.

\subsection{First order solution of \eqref{pdeSeries}} \label{firstOrder}

To construct the solution in the first order, we keep first two terms in \eqref{series1}, and ignore all terms $O(\xi^2)$. Thus, in this approximation $z(u;t,\sigma,\calV) = z_0(u;t,\sigma,\calV) + \xi z_1(u;t,\sigma,\calV)$. Then from \eqref{pdeSeries} we obtain
\begin{align} \label{pdeSeries1}
0 &=  \fp{z_1}{t} + {\cal L} z_1 + \varPhi_1 z_0, \\
\varPhi_1 &= \nu^2(t) \sigma \cp{}{\calV}{\sigma}
+ \left[m(t) \calV + \iu u \rho \nu(t) \sigma\right] \sigma \fp{}{\sigma}. \nonumber
\end{align}
Thus, this equation acquires almost the same form as \eqref{pdeSeries0}, but with two important changes. First, it has an additional source term $\varPhi_1 z_0$. Second, as the terminal condition $z(u; T, \sigma, \calV) = 1$ is already satisfied by the zero-order approximation $z_0(u;t,s,\calV)$, this equation has to be solved subject to the vanishing initial condition $z_1(u; T, \sigma, \calV) = 0$.

It is well known from the theory of PDEs, e.g., see \cite{Polyanin2002}, that the general solution of \eqref{pdeSeries1} can be represented as
\begin{align} \label{source}
z_1(u; t, \sigma, \calV) &= \int_{-\infty}^\infty \int_{0}^\infty
z_1(u; T, \sigma', \calV') \calG_1(\sigma,\sigma',\calV,\calV',T-t) d\sigma' d\calV' \\
&- \int_T^t \int_{-\infty}^\infty \int_{0}^\infty \varPhi(u; k, \sigma', \calV')  \calG_1(\sigma,\sigma',\calV, \calV',k-t) d \sigma' d\calV' dk, \nonumber \\
\varPhi(u; k, \sigma', \calV') &= \varPhi_1 z_0(u; k, \sigma', \calV'), \nonumber
\end{align}
\noindent where $z_1(u; T, \sigma', \calV')$ is the terminal condition, and  $\calG_1(\sigma,\sigma',\calV,\calV',t)$ is the Green function of the homogeneous counterpart of \eqref{pdeSeries1}. Since in our case $z_1(u; T, \sigma', \calV') = 0$, the first integral in \eqref{pdeSeries1} disappears.
Also, as the homogeneous counterpart of \eqref{pdeSeries1} has exactly same structure as \eqref{pdeSeries0}, the Green function  $\calG(\sigma,\sigma',\calV,\calV',t)$ can be transformed to those found in Section~\ref{GRF}.

In more detail, computation of the second integral in \eqref{source} can be done as follows. First, we re-write it as
\begin{equation} \label{I0}
{\cal I}_1 = \varPhi_1 {\cal I}, \quad
{\cal I} = \int_t^T \int_{-\infty}^\infty \int_{0}^\infty z_0(u; k, \sigma', \calV')  \calG(\sigma,\sigma',\calV, \calV',k-t) d \sigma' d\calV' dk.
\end{equation}
Second, we represent $z_0(u; t, \sigma, \calV)$ in variables $w_1(u;\tau(t), \omega), w_2(u;\tau, \varsigma)$
using a series of transformations presented in Section~\ref{zeroOrder}
\begin{align}
z_0(u;t,\sigma,\calV) &=  \exp\left[a(t) + \gamma(t) \sigma^2 + \beta(t) \sigma \calV \right]
y_0(u;t,\sigma,\calV) \\
&=   \exp\left[a(t) + \gamma(t) \left(\frac{g}{\calV}\right)^2 + \beta(t) g \right]
y_0(u;t,g,\calV) \nonumber \\
&= \exp\left[a(t) + \omega^2 e^{-2\kappa t}\frac{\gamma(t)  \alpha^2_1(t)}{[\varsigma - 2 \tau(t)]^2} + \beta(t) e^{-\kappa t} \omega  + \varsigma - \tau(t) \right]  w_1(u;\tau(t), \omega) w_2(u;\tau(t), \varsigma) \nonumber \\
&= \exp\left[a(t) - \tau(t) + \omega^2 e^{-2\kappa t}\frac{ \gamma(t)  \alpha^2_1(t)}{[\varsigma - 2 \tau(t)]^2} + \bar{\beta}(t) e^{-\kappa t} \omega  +  \varsigma  \right]. \nonumber
\end{align}

Then using the Green functions found in \eqref{w1G}, \eqref{w2G}, we obtain
\begin{align}
{\cal I} = \int_0^\tau \int_{-\infty}^\infty
&\exp \Bigg[ a(t) - \chi + \omega^2 e^{-2 \kappa t}\frac{\gamma(t)  \alpha^2_1(t)}{(\varsigma' - 2 \chi)^2} +
f(t) \omega + \varsigma'  - \frac{(\varsigma - \varsigma')^2}{4 (\chi-\tau)} \Bigg] G(\chi)  \\
 &\cdot \frac{ 1 - \Theta(\tau-\chi) + \Theta(0)}{2\sqrt{\pi (\chi-\tau)}}   d \varsigma' d \chi, \nonumber \\
 f(t) &= \bar{\beta}(t) e^{-\kappa t}  +  \int_0^{\chi-\tau} a_1(t(k)) G(t(k)) dk, \qquad G(t) = \frac{1}{d \tau(t)/dt}, \nonumber
\end{align}
\noindent where $t = t(\chi)$ and $t = t(k)$ are the inverse of the function $\tau(t)$. Also, according to \eqref{Htrans},
\[ \fp{\tau(t)}{t} = - \frac{1}{2} \nu^2(t) \alpha_1^2(t). \]

Switching back from $\tau$ to $t$, we obtain
\begin{align} \label{I1fin}
{\cal I} &= \int_t^T \frac{ 1 - \Theta(t-\chi) + \Theta(0)}{2\sqrt{\pi (\chi-t)}} e^{a(\chi) - \chi + f_1(\chi) \omega}
{\cal J}(t,\varsigma, \omega; \chi) d \chi,  \\
{\cal J}(t,\varsigma, \omega; \chi) &= \int_{-\infty}^\infty  \exp\left[ \omega^2 e^{-2\kappa \chi}
 \frac{\gamma(\chi) \alpha^2_1(\chi)}{(\varsigma' - 2 \chi)^2} + \varsigma'   - \frac{(\varsigma' - \varsigma)^2}{4 (\chi-t)} \right] d \varsigma', \nonumber \\
f_1(\chi) &= \bar{\beta}(\chi) e^{-\kappa \chi}  +  \int_0^{\chi-t} a_1(k) dk. \nonumber
\end{align}
As follows from the definition of $\gamma(\chi)$ in \eqref{st1}, $\gamma(\chi) \le 0$. Therefore, the second integral in  \eqref{I1fin} is well-defined.

Finally, to obtain $z_1(u; t, \sigma, \calV)$, in \eqref{I1fin} we set $t=0$, substitute $\omega = e^{\kappa\chi} \sigma \calV, \ \varsigma = \alpha_1(\chi) \calV + 2 \tau(\chi)$, and apply operator $\varPhi_1$ to the result.

\subsection{Second order solution of \eqref{pdeSeries}} \label{secondOrder}

In the second order approximation on $\xi$ \eqref{pdeSeries} transforms to
\begin{align} \label{pdeSeries2}
0 &=  \fp{z_2}{t} + {\cal L} z_2 + \varPhi_1 z_1 + \varPhi_2 z_0, \\
\varPhi_2 &= \frac{1}{2}\nu^2(t) \sigma^2 \sop{}{\sigma}.  \nonumber
\end{align}
Again, this equation acquires the same form as \eqref{pdeSeries1}, but with a slightly different source term. Also, similar to \eqref{pdeSeries1}, as the terminal condition $z(u; T, \sigma, \calV) = 1$ is already satisfied by the zero-order approximation $z_0(u;t,s,\calV)$, \eqref{pdeSeries2} should be solved subject to the vanishing terminal condition $z_2(u; T, \sigma, \calV) = 0$.

Thus, the solution at this step is given by \eqref{source} with
\begin{equation} \label{source1}
\varPhi(u; k, \sigma', \calV') = \varPhi_2 z_0(u; k, \sigma', \calV') +
\varPhi_1 z_1(u; k, \sigma', \calV'),
\end{equation}
\noindent where the first integral in \eqref{source} again vanishes due to the terminal condition. The second integral can be computed in the same way as this was done for the first-order approximation, as the the Green function of the homogeneous PDe is already known.

In principle, the higher order approximations could be constructed in a similar way, as the Green function doesn't change, but only the source term. The higher order PDEs take the form
\begin{equation}
0 =  \fp{z_i}{t} + {\cal L} z_i + \varPhi_1 z_{i-1} + \varPhi_2 z_{i-2}, \qquad i > 1,
\end{equation}
\noindent and should be solved subject to the vanishing terminal condition $z_i(u; T, \sigma, \calV) = 1$. Again, the solution is given by \eqref{source} with
\begin{equation} \label{source2}
\varPhi(u; k, \sigma', \calV') = \varPhi_2 z_{i-2}(u; k, \sigma', \calV') +
\varPhi_1 z_{i-1}(u; k, \sigma', \calV').
\end{equation}

\section{An example}

Looking closely at the integrand of ${\cal J}(0,\varsigma, \omega; \chi)$ in \eqref{I1fin}, one can observe that it behaves as follows:
\begin{enumerate}
\item Suppose we consider options with maturities $T < 1$  year. Since $0 \le \chi \le T$, at $\varsigma'$ far away from $\varsigma$  the term $\frac{(\varsigma' - \varsigma)^2}{4 (\chi-t)}$ is large. Therefore, for these regions of $\varsigma'$ the integrand almost vanishes.

\item Also, at large $|\varsigma'|$ where $(\varsigma' - 2x)^2$  is also large, the term
\[ \omega^2 e^{-2\kappa \chi}  \frac{\gamma(\chi) \alpha^2_1(\chi)}{(\varsigma' - 2 \chi)^2} =
 -\sigma^2 \calV^2 \frac{u[1 + u (1 - \rho)^2]}{4 \kappa} \left(1 - e^{2 \kappa (\chi-T)} \right).
\frac{ \alpha^2_1(\chi)}{(\varsigma' - 2 \chi)2}
\]
is small for $\sigma, \calV, u$ fixed.
\end{enumerate}

Thus, the integrand of ${\cal J}(0,\varsigma, \omega; \chi)$ has a bell shape with a maximum close to the point $\varsigma' = \varsigma_*$ which solves the equation
\begin{equation} \label{x*}
\partial_{\varsigma'} \left[\omega^2 e^{-2\kappa \chi}
 \frac{\gamma(\chi) \alpha^2_1(\chi)}{(\varsigma' - 2 \chi)^2} + \varsigma'   - \frac{(\varsigma' - \varsigma)^2}{4 (\chi-t)}\right] = 0.
 \end{equation}

 Indeed,  consider an example with the explicit form of the function $m(t) = \varrho t^\pi, \ \varrho, \pi \in \mathbb{R}, \ \pi \ge 0$, so the SDE for $\calV_t$ in \eqref{ADOhestonQ} is mean-reverting. With this $m(t)$ one can find that
\begin{equation} \label{gammaExp}
\gamma(t) = \frac{B_H^2 }{2(1+\pi)} e^{- \frac{\varrho}{1+\pi} T^{1+\pi}}
\left[t^{2H} E\left(1-\frac{2 H}{1+\pi}, - \frac{\varrho t^{1+\pi}}{1+\pi}  \right) -
T^{2H} E\left(1-\frac{2 H}{1+\pi}, - \frac{\varrho T^{1+\pi}}{1+\pi} \right) \right],
\end{equation}
\noindent where $E(k, z)$ is the exponential integral function, \cite{as64}. Let's also use the values of  our model parameters given in Table~\ref{param}.

  \begin{table}[!htb]
\begin{center}
\begin{tabular}{|c|c|c|c|c|c|c|c|c|}
\hline
$\kappa$ & $H$ & $T$ & $\sigma$ & $\calV$ & $\rho$ & $u$ & $\varrho$ & $ \pi$  \\
\hline
2.0 & 0.3 & 0.5 & 0.3 & 5.0 & -0.5 & 1.0 & 1.0 & 0.5 \\
\hline
\end{tabular}
\caption{Parameters of the test.}
\label{param}
\end{center}
\end{table}

\begin{figure}[!htb]
\begin{minipage}{0.4\textwidth}
\begin{center}
%\onelinecaptionsfalse
\includegraphics[totalheight=2.4in]{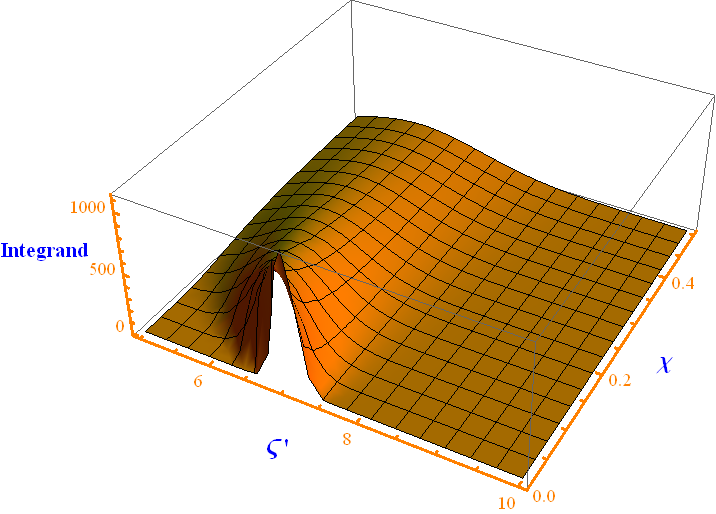}
\caption{Integrand of ${\cal J}(0,\varsigma, \omega; \chi)$ as function of $(\varsigma', \chi)$ at $0 \le \chi \le T$.}
\label{bigScale}
\end{center}
\end{minipage}
\hspace{0.1\textwidth}
\begin{minipage}{0.4\textwidth}
\begin{center}
%\onelinecaptionsfalse
\includegraphics[totalheight=2.4in]{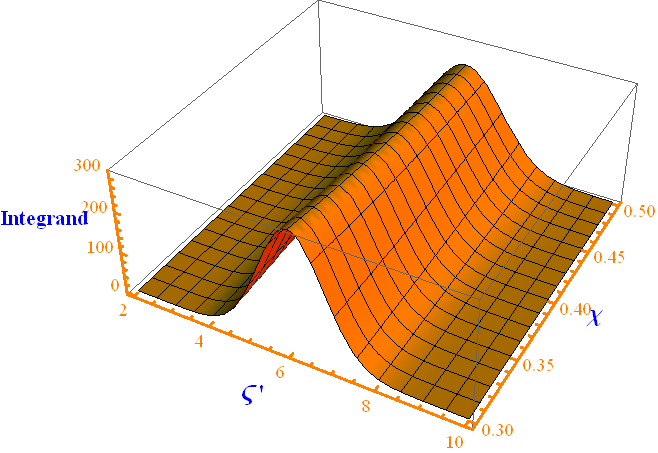}
\caption{Integrand of ${\cal J}(0,\varsigma, \omega; \chi)$ as function of $(\varsigma', \chi)$ at $0.3 \le \chi \le T$.}
\label{smallScale}
\end{center}
\end{minipage}
\end{figure}

 Now the integrand of ${\cal J}(0,\varsigma, \omega; \chi)$ can be computed explicitly, and the result is presented in Fig.~\ref{bigScale}. The bell shape of this function could be clearly seen at small $T$. To make sure a similar shape could be seen at large $T$, we zoom-in this plot in $\chi$, and the result is presented in Fig.~\ref{smallScale} which justifies the previous observation.

The bell shape of the integrand implies that the integral ${\cal J}(0,\varsigma, \omega; \chi)$ an be computed approximately in closed form. Indeed, the maximum of the integrand approximately corresponds to the point  $\varsigma' = \varsigma$ where $\varsigma = \alpha_1(\chi) \calV + 2 \tau(\chi)$. Then, using the representation of the integrand in the form
\begin{align}
f(\chi, \varsigma, \varsigma') &= \frac{k(\chi)}{(\varsigma' - 2 \chi)^2} + \varsigma'   - \frac{(\varsigma' - \varsigma)^2}{4 \chi}, \\
k(\chi) &=  \gamma(\chi) \alpha^2_1(\chi) \sigma \calV, \nonumber
\end{align}
\noindent we expand $f(\chi, \varsigma, \varsigma')$ into series on $\varsigma'$ around $\varsigma$ to obtain
\begin{align} \label{quadExp}
f(\chi, \varsigma, \varsigma') &= a_0 + a_1 (\varsigma'-\varsigma) + a_2 (\varsigma'-\varsigma)^2 + O((\varsigma'-\varsigma)^3), \\
a_0 &= \varsigma + \frac{k(\chi)}{(\varsigma - 2 \chi)^2}, \quad
a_1 = 1 - \frac{2 k(\chi)}{(\varsigma - 2 \chi)^3}, \quad
a_2 = - \frac{1}{4 \chi} + \frac{3 k(\chi)}{(\varsigma - 2 \chi)^4}. \nonumber
\end{align}

Then
\begin{equation} \label{appr}
{\cal J}(0,\varsigma, \omega; \chi) = \sqrt{\frac{\pi}{-a_2}} e^{ a_0 - \frac{a_1^2}{4 a_2}},
\end{equation}
\noindent which exists if $a_2 < 0$.

\begin{figure}[!htb]
\begin{minipage}{0.4\textwidth}
\begin{center}
%\onelinecaptionsfalse
\includegraphics[totalheight=2.in]{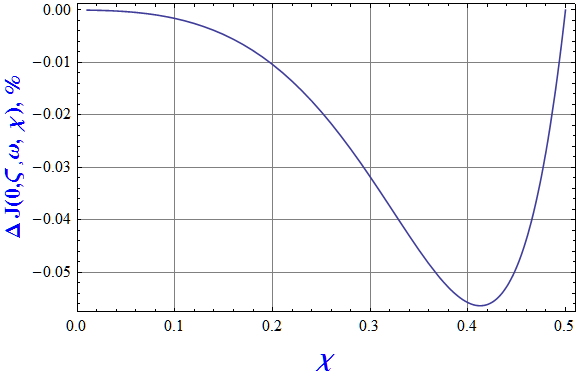}
\caption{Difference in \% between ${\cal J}(0, \varsigma, \omega; \chi)$ obtained by using numerical integration and \eqref{appr}.}
\label{difApp}
\end{center}
\end{minipage}
\hspace{0.1\textwidth}
\begin{minipage}{0.4\textwidth}
\begin{center}
%\onelinecaptionsfalse
\includegraphics[totalheight=2.in]{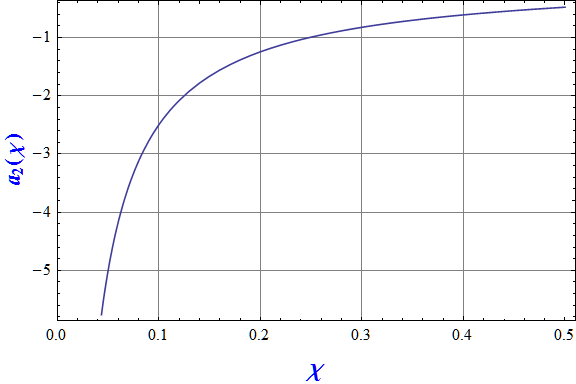}
\caption{Function $a_2(\chi)$ computed for our experimant. \newline \newline}
\label{a2chi}
\end{center}
\end{minipage}
\end{figure}

Fig.~\ref{a2chi} demonstrates function $a_2(\chi)$ computed in this experiment which turns out to be negative for all values of $0 \le \chi \le T$. Then Fig.~\ref{difApp} presents a percentage difference between the value of ${\cal J}(0, \varsigma, \omega; \chi)$ computed numerically and using \eqref{appr}. The difference is about 5 bps, so in out test
this aproximation works pretty well.

Alternatively, \eqref{x*} is a quartic algebraic equation which can be solved in closed form. Denoting this solution by $\varsigma_*$ and using it instead of $\varsigma$ in \eqref{quadExp}, we obtain another approximation.

\section{Discussion}

As the CF of the $\log S_T$ is known in closed form (in our case this is an approximation of the exact solution constructed by using power series in $\xi$), pricing options can be done in a standard way by using FFT, \cite{CarrMadan:99a, Lewis:2000, FangOosterlee2008}. In turn, pricing variance swaps can be done by using a forward CF, similar to how this is done in \cite{ItkinCarr2010}. Using the forward time $t$ the forward characteristic function is defined as
\begin{equation}\label{fcf}
\phi_{t,T} = \EQ \left[\exp (\iu u \calx_{t,T}) | S_0, \sigma _0 \right],
\end{equation}
\noindent where $\calx_{t, T} = \calx_T - \calx_t$, and $\calx_t = \log S_t$. Then under a discrete set of observations of the stock price at times $t_i, \ i \in [1,N]$, the quadratic variation  $\mathcal{Q}_N(x)$ of $S_t$ is given by, \cite{ItkinCarr2010}
\begin{align}\label{qv3}
 \mathcal{Q}_N(s) &= \dfrac{1}{T}\sum_{i=1}^N \EQ \left[ (\calx_{t_i} - \calx_{t_{i-1}})^2\right]
  = \dfrac{1}{T}\sum_{i=1}^N \EQ \left[ \calx_{ t_i, t_{i-1} }^2\right] = - \dfrac{1}{T}\sum_{i=1}^N \dfrac{\partial^2 \phi_{ t_i, t_{i-1} }(u) }{\partial u^2}\Big|_{u=0}.
\end{align}
As $\calx_{t_i, t_{i-1}} = \log S_{t_i} - \log S_{t_{i-1}} = x_{T-t_{i-1}} - x_{T-t_{i}}$, $\mathcal{Q}_N(s)$ in \eqref{qv3} can be computed in a way similar to how this was done in Section~\ref{CFT}.

\begin{figure}[!htb]
\begin{minipage}{0.4\textwidth}
\begin{center}``
%\onelinecaptionsfalse
\includegraphics[totalheight=2.5in]{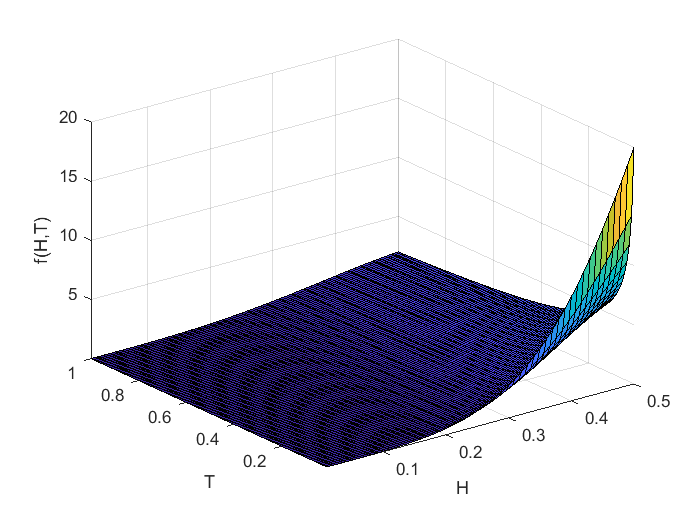}
\caption{Function $f(H,T)$ for $H \in [0,1]$ and $T \in [0,1]$, years}
\label{smalPar}
\end{center}
\end{minipage}
\hspace{0.1\textwidth}
\begin{minipage}{0.4\textwidth}
\begin{center}
%\onelinecaptionsfalse
%\DeclareGraphicsRule{.jpg}{eps}{.bb}{}% declare GIF filename extension
\includegraphics[totalheight=2.5in]{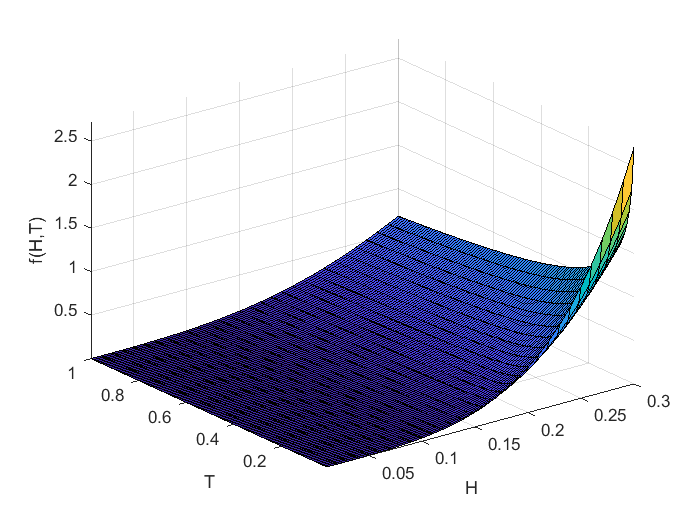}
\caption{Function $f(H,T)$ for $H \in [0,0.3]$ and $T \in [0,1]$, years}
\label{smalParShort}
\end{center}
\end{minipage}
\end{figure}

Therefore, the proposed model could be useful, e.g., for pricing options and swaps as, on the one hand, it catches some properties of rough volatility, but, on the other hand, is more tractable. We underline, that our approach allows the CF to be found as the solution of the PDE  in \eqref{pde11}. This PDE, in general, can be solved numerically. But in this paper we provide a closed-form series solution obtained by assuming the vol-of-vol $\xi$ to be small.  This condition is defined in \eqref{smallVV}, so, as can be seen, it is a function of the HUrst exponent $H$ and time to maturity $T$. The function $f(H,T) = 2/(B_H T^H)$ for various values of $H$ and $T$ is represented in Fig.~\ref{smalPar}. As the range $H \in [0, 0.3]$ is reported in the literature to be important, we re-plot this graph in Fig.~\ref{smalParShort} by zooming into this area.

Overall, the values of $f(H,T)$ look reasonable as compared with those reported in the literature, for instance, for the Heston model. In other words, the values of the vol-of-vol parameter $\xi$, found by calibration of the Heston model to market prices of European vanilla options, could be of the order of magnitude to obey \eqref{smallVV} for $H > 0.1$ and $T > 0.1$. However, this definitely should be justified by independent calibration of the ADOL model to those market data.
This calibration would require solving the PDE in \eqref{pde11} numerically, to not rely on the assumption \eqref{smallVV}.  These results will be reported elsewhere.

Another assumption we made when deriving \eqref{ADOheston3} is that the mean-reversion level $\theta=0$. With a little algebra it can be checked, that relaxing this assumption adds an extra term to \eqref{pdePsi1}  which is $\kappa \theta\fp{z_0}{\sigma}$. Accordingly, in the definition of operator $\cal L$ in \eqref{pdeSeries} this also adds an extra term $\kappa \theta \fp{}{\sigma}$. The next step is to make a change of variable $\sigma \mapsto \sigma - \theta$. As $\theta$ is assumed to be constant, \eqref{pdeSeries0-1} could again be replicated if in \eqref{st1} we add an extra term $\bar{\gamma}(t) s$, i.e.
\begin{align} \label{st21}
z_0(u;t,\sigma,\calV) &\mapsto y_0(u;t,\sigma,\calV) \exp\left[a(t) + \bar{\gamma}(t) s + \gamma(t) \sigma^2 + \beta(t) \sigma \calV \right], \\
\bar{\gamma}(t) &= \frac{\theta  u (u+1)}{\kappa} \left(e^{k (t-T)} - 1\right). \nonumber
\end{align}
Then construction of the solution remains the same.

%%%%%%%%%%%%%%%%%%%%%%%%%%%%%%%%%%%%%%%%%%%%%%%%%%%%%%%%%%%%
\vspace{0.3in}
%\printbibliography[title={Bibliography}]

\newcommand{\noopsort}[1]{} \newcommand{\printfirst}[2]{#1}
  \newcommand{\singleletter}[1]{#1} \newcommand{\switchargs}[2]{#2#1}


\begin{thebibliography}{}

\bibitem[Abramowitz and Stegun, 1964]{as64}
Abramowitz, M. and Stegun, I. (1964).
\newblock {\em Handbook of Mathematical Functions}.
\newblock Dover Publications, Inc.

\bibitem[Benhamou et~al., 2010]{Benhamou2010}
Benhamou, E., Gobet, E., and Miri, M. (2010).
\newblock Time dependent heston model.
\newblock {\em {SIAM} Jounral of Financial Mathematics}, 1:289--325.

\bibitem[Benth and Khedher, 2016]{StochMR2016}
Benth, F. and Khedher, A. (2016).
\newblock {Weak Stationarity of Ornstein-Uhlenbeck Processes with Stochastic
  Speed of Mean Reversion}.
\newblock In Podolskij, M., Stelzer, R., Thorbjornsen, S., and Veraart, A.,
  editors, {\em The Fascination of Probability, Statistics and their
  Applications}. Springer, Cham.

\bibitem[Bi et~al., 2016]{Bi2016}
Bi, M., Escobar, M., Goetz, B., and Zagst, R. (2016).
\newblock Principal component models with stochastic mean reverting levels.
  pricing and covariance surface improvements.
\newblock {\em Applied Stochastic models in Business and Industry}.

\bibitem[Carr and Madan, 1999]{CarrMadan:99a}
Carr, P. and Madan, D. (1999).
\newblock Option valuation using the {Fast Fourier Transform}.
\newblock {\em Journal of Computational Finance}, 2(4):61--73.

\bibitem[Carr and Wu, 2004]{CarrWu2004}
Carr, P. and Wu, L. (2004).
\newblock Time-changed levy processes and option pricing.
\newblock {\em Journal of Financial Economics}, 71(1):113--141.

\bibitem[Christoffersen et~al., 2010]{CJM2010}
Christoffersen, P., Jacobs, K., and Mimouni, K. (2010).
\newblock Models for s\&p 500 dynamics: Evidence from realized volatility,
  daily returns and options prices.
\newblock {\em Review of Financial Studies}, 23(9):3141--3189.

\bibitem[Conus and Wildman, 2016]{Conus2016}
Conus, D. and Wildman, M. (2016).
\newblock A gaussian markov alternative to fractional brownian motion for
  pricing financial derivatives.
\newblock available at arXiv:1608.03428v1.

\bibitem[Dobri\'{c} and Ojeda, 2006]{DObook2006}
Dobri\'{c}, V. and Ojeda, F.~M. (2006).
\newblock {\em Fractional Brownian fields, duality, and martingales}.
\newblock Institute of Mathematical Statistics Lecture Notes - Monograph
  Series. Institute of Mathematical Statistics, Beachwood, Ohio, USA.

\bibitem[Dobri\'{c} and Ojeda, 2009]{DobricOjeda2009}
Dobri\'{c}, V. and Ojeda, F.~M. (2009).
\newblock Conditional expectations and martingales in the fractional brownian
  field.
\newblock In {\em Institute of Mathematical Statistics Collections}, pages
  224--238.

\bibitem[{El\ Euch} and Rosenbaum, 2016]{EuchRos2016}
{El\ Euch}, O. and Rosenbaum, M. (2016).
\newblock The characteristic function of rough heston models.
\newblock available at \url{https://arxiv.org/pdf/1609.02108.pdf}.

\bibitem[Fang and Oosterlee, 2008]{FangOosterlee2008}
Fang, F. and Oosterlee, C. (2008).
\newblock A novel pricing method for {E}uropean options based on
  {Fourier-Cosine} series expansions.
\newblock {\em SIAM J Sci Comput}, 31(2):826--848.

\bibitem[Funahashi and Kijima, 2017]{Funahashi2017}
Funahashi, H. and Kijima, M. (2017).
\newblock A solution to the time-scale fractional puzzle in the implied
  volatility.
\newblock {\em Fractal and fractional}, 1(1):14--31.

\bibitem[Gatheral, 2006]{Gatheral2006}
Gatheral, J. (2006).
\newblock {\em The volatility surface}.
\newblock Wiley finance.

\bibitem[Gatheral, 2008]{Gatheral2008}
Gatheral, J. (2008).
\newblock Consistent modeling of {SPX} and {VIX} options.
\newblock In {\em {Fifth World Congress of the Bachelier Finance Society}}.

\bibitem[Gatheral et~al., 2014]{GatheralJaissonRos2014}
Gatheral, J., Jaisson, T., and Rosenbaum, M. (2014).
\newblock Volatility is rough.
\newblock Available at SSRN 2509457.

\bibitem[Guennoun et~al., 2014]{GueJacRoome2014}
Guennoun, H., Jacquier, A., and Roome, P. (2014).
\newblock Asymptotic behaviour of the fractional heston model.
\newblock Available at SSRN 2531468.

\bibitem[Hagan et~al., 2002]{hagan2002}
Hagan, P., Kumar, D., A, A.~L., and Woodward, D. (2002).
\newblock Managing smile risk.
\newblock {\em Wilmott magazine}, pages 84--108.

\bibitem[Harms, 2019]{Harms2019}
Harms, P. (2019).
\newblock {Strong convergence rates for Markovian representations of fractional
  Brownian motion}.
\newblock arXiv: 1902.02471.

\bibitem[Heston, 1993]{Heston:93}
Heston, S. (1993).
\newblock Closed-form solution for options with stochastic volatility, with
  applicationto bond and currency options.
\newblock {\em Review of Financial Studies}, 6(2):327--343.

\bibitem[Itkin and Carr, 2010]{ItkinCarr2010}
Itkin, A. and Carr, P. (2010).
\newblock Pricing swaps and options on quadratic variation under stochastic
  time change model - a discrete observation case.
\newblock {\em Review Derivatives Research}, 13:141--176.

\bibitem[Karatzas and Shreve, 1991]{karatzas1991brownian}
Karatzas, I. and Shreve, S. (1991).
\newblock {\em {Brownian Motion and Stochastic Calculus}}.
\newblock Graduate Texts in Mathematics. Springer, New York.

\bibitem[Lewis, 2000]{Lewis:2000}
Lewis, A.~L. (2000).
\newblock {\em Option Valuation under Stochastic Volatility}.
\newblock Finance Press, Newport Beach, California, USA.

\bibitem[Livieri et~al., 2018]{Livieri2018}
Livieri, G., Mouti, S., Pallavicini, A., and Rosenbaum, M. (2018).
\newblock Rough volatility: Evidence from option prices.
\newblock {\em IISE Transactions}, 50(9):767--776.

\bibitem[Lord et~al., 2007]{LordAmerican2007}
Lord, R., Fang, F., Bervoets, F., and Oosterlee, C. (2007).
\newblock A fast and accurate fft-based method for pricing early-exercise
  options under levy processes.
\newblock SSRN: 966046.

\bibitem[Polyanin, 2002]{Polyanin2002}
Polyanin, A. (2002).
\newblock {\em Handbook of linear partial differential equations for engineers
  and scientists}.
\newblock Chapman \& Hall/CRC.

\bibitem[Rouah, 2013]{Rouah2013}
Rouah, F. (2013).
\newblock {\em Heston model and its extensions in Matlab and C\#}.
\newblock John Wiley \& Sons, Inc.,, Hoboken, New Jersey.

\bibitem[Sepp, 2016]{Sepp2016}
Sepp, A. (2016).
\newblock Log-normal stochastic volatility model: Affine decomposition of
  moment generating function and pricing of vanilla options.
\newblock SSRN-2522425.

\bibitem[Shreve, 1992]{Shreve:1992}
Shreve, S. (1992).
\newblock Martingales and the theory of capital-asset pricing.
\newblock {\em Lecture Notes in Control and Information SCIENCES},
  180:809--823.

\bibitem[Wildman, 2016]{Wildman2016}
Wildman, M. (2016).
\newblock {\em The Dobric-Ojeda Process with Applications to Option Pricing and
  the Stochastic Heat Equation}.
\newblock PhD thesis, Lehigh University.

\bibitem[Wong and Heyde, 2006]{WongHeyde2006}
Wong, B. and Heyde, C.~C. (2006).
\newblock On changes of measure in stochastic volatility models.
\newblock {\em Journal of Applied Mathematics and Stochastic Analysis}, (ID
  18130).

\end{thebibliography}
\end{document}